# Nonlinear oscillatory mixing in the generalized Landau scenario

R. Herrero [a], J. Farjas [b], F. Pi [c], G. Orriols [c]

[a] Departament de Física i Enginyeria Nuclear, Universitat Politècnica de Catalunya, 08222 Terrassa, Spain.

[b] Departament de Física, Campus Montilivi, Universitat de Girona, 17003 Girona, Spain.

[c] Departament de Física, Universitat Autònoma de Barcelona, 08193 Cerdanyola del Vallès, Spain.

We present a set of phase-space portraits illustrating the extraordinary oscillatory possibilities of the dynamical systems through the so-called generalized Landau scenario. In its simplest form the scenario develops in $N$ dimensions around a saddle-node pair of fixed points experiencing successive Hopf bifurcations up to exhausting their stable manifolds and generating $N$-1 different limit cycles. The oscillation modes associated with these cycles extend over a wide phase-space region by mixing ones within the others and by affecting both the transient trajectories and the periodic orbits themselves. A mathematical theory covering the mode-mixing mechanisms is lacking, and our aim is to provide an overview of their main qualitative features in order to stimulate research on it.

## I. Introduction

The intuitively convincing idea that complex oscillations could be achieved by combining more and more oscillations, as proposed by Landau to tentatively explain the transition to turbulence [1], has not found a definite way in the mainstream of nonlinear dynamics. The Landau's proposal was based on a succession of two-dimensional instabilities generating quasiperiodic states with successively additional frequencies but it was shown [2,3] that, after a few of such instabilities, the underlying invariant torus loses smoothness and breaks down, usually leading to the occurrence of chaos. As a matter of fact, invariant tori of order higher than two have been rarely observed in autonomous dissipative systems [4-6] and, in the literature, the term of complex oscillations usually appears in reference to the irregularities of chaos. The term is also used in relation to the so-called mixed-mode oscillations in which there is an alternation between oscillations of distinct large and small amplitudes [7]. Mixed-mode oscillations are typically observed in fast-slow systems involving local phase-space phenomena like folded singularities, canard orbits or singular Hopf bifurcations, through which the structure in the oscillatory behavior is originated. With such a kind of mechanisms, structured periodic oscillations with three components of different amplitudes and frequencies have been numerically obtained in two feedforward coupled FitzHugh-Nagumo systems [8] and in a four-dimensional piecewise linear system [9]. However, the combination of such a kind of mechanisms to provide a generic way for the achievement of really complex oscillations with arbitrarily large numbers of oscillatory modes seems not obvious. To the best of our knowledge, the unique known way for such a purpose is the so-called generalized Landau scenario, which is based on the reiterative occurrence of the two most standard mechanisms of nonlinear dynamics: the saddle-node and Hopf bifurcations, and which, despite having been the object of several publications [10-12], remains unnoticed in the field. This article is another attempt in the form of a graphical exposition. In relation to previous works, it provides a more comprehensive overview of the oscillatory scenario by illustrating how the several oscillation modes extend for the phase space and how ones mix with others by affecting both the transient trajectories and the periodic orbits themselves. Particularly relevant is the bidirectional mixing of modes between those associated with the node and those with the saddle fixed point. Also relevant is the fact that, along with the frequency, each oscillation mode appears with the same phase-space orientation everywhere its influence is manifested, denoting a well-defined association of the mode with a given dynamical activity of the system variables.The geometrical view also indicates that there is no reason for a limit in the number of mixed modes other than the phase space dimension and that the scenario might be truly scalable.

## II. Oscillatory mixing scenario

We are not going to repeat the contents of previous works, where the oscillatory behavior is widely illustrated through time evolution signals of the attractor and where some basic ingredients of the scenario unfolding are analyzed [12] (for a descriptive overview see Appendix 3 of Ref. [13]). We here restrict our comments to describing the phase portraits and to remarking on the variety of qualitative features to be covered by any tentative theoretical analysis of the oscillatory mixing scenario.

The numerical demonstrations have been done with the $N$-dimensional system

$$\frac{dz_1}{dt} = -\sum_{q=1}^{N} c_q z_q + \mu_C g(\psi), \qquad \frac{dz_j}{dt} = z_{j-1}, \; j = 2, \ldots, N, \tag{1}$$

with

$$\psi = \sum_{q=1}^{N} d_q z_q, \tag{2}$$

and where the components of the $N$-dimensional vector $z$ are the system variables, $g(\psi)$ denotes a relatively arbitrary nonlinear function of a single variable that in its turn is a linear combination of all the system variables, the scale-factor $\mu_C$ is used as a control parameter, and the values for the two sets of $c_q$ and $d_q$ coefficients have to be properly determined such that the $\mu_C$-family of systems will develop the oscillatory scenario with increasing $\mu_C$. The system (1) was introduced and analyzed in Ref. [10], as a generalization of a model describing the experimental behaviors of a family of physical devices with successively increasing dimension (up to 6) [11], and it is succinctly presented in the Appendix, together with a consideration of what of its features are responsible for the good working of the oscillatory scenario.

As it is described in the Appendix, the key point of system (1) is its peculiarity of admitting to be easily designed such that it will possess a saddle-node pair of fixed points experiencing successive Hopf bifurcations up to exhausting their stable manifolds, with a total number of *N*-1 bifurcations and with prechosen values for the corresponding frequencies. If properly done, the design procedure provides $\mu_C$-families of *N*-dimensional systems exhibiting the nonlinear mixing of *N*-1 oscillation modes in a generalized Landau scenario. More complex scenarios involving large sets of fixed points and generating large numbers of oscillation modes are also possible but they require additional nonlinearities, i.e., systems like Eqs. (A1) with $m > 1$, and will not be considered here.

The reported simulations correspond to the $\mu_C$-families of systems with $N = 4$ and $N = 6$ whose values for the $c_q$ and $d_q$ coefficients are given in Table A1 and with the nonlinear function $g(\psi)$ defined by either Eq. (A7) or Eq. (A8) and denoted $g_A$ or $g_B$, respectively. Concretely, the illustrations deal with three different families of systems: $N = 4$ with $g_A$ (Fig. 1), $N = 6$ with $g_A$ (Figs. 2 to 7) and $N = 6$ with $g_B$ (Figs. 8 and 9).

As it is illustrated in Fig. 1 for $N = 4$, the scenario unfolding as a function of $\mu_C$ develops through the gradual appearance of the various oscillation modes in certain phase space regions, so that the transient trajectories crossing these regions successively manifest the corresponding oscillations, and through the appearance of fixed points and of limit cycles from these points. Notice how the various oscillation modes and their mixing manifest on the trajectories even before the appearance of the corresponding periodic orbits and even before the appearance of the fixed points from which the periodic orbits will emerge. Notice also that the periodic orbit W1 experiences the oscillatory mixing by incorporating influences of the other two modes and, most important, that this happens without necessity of doing any bifurcation. As will be shown below, the incorporation of other modes in a periodic orbit without its bifurcation is a generic feature of the scenario. On the whole, the numerical observations strongly suggest the global nature of the oscillatory scenario and the topological implications of its development through the nonlinear mode-mixing mechanisms.

Our previous publications [10-12] deal with a limited view of the oscillatory scenario, essentially based on the observation of the attractor W1, and our interpretative trials were mainly devoted to explain the influence of the rest of modes on this attractor. The tentative explanation was based on the combination of two basic mechanisms of mode mixing between pairs of periodic orbits (see, e.g., Figs 1 and 2 of Ref. [12]): on the one hand, the mixing between modes emerged from the same fixed point, either the node or the saddle, and, on the other hand, the influence of the modes of the saddle towards those of the node. The working of both kinds of mixing mechanisms was sustained through the unstable submanifold connection from the influencing periodic orbit to the influenced one. For instance, in Fig. 1(d) the three-dimensional unstable manifold of W3 (not drawn in the figure) goes fully towards the attractor W1 and would transmit the mode 3 oscillations towards it. The mode-mixing influences appear always localized and, in this case, they affect two different places of W1. In Fig. 1(e), the two-dimensional unstable manifold of W2 goes down towards the attractor W1 by transmitting the mode 2 oscillations like a sort of corkscrew and affecting the crossed phase-space region also. In this case the mode 2 appears also localized on W1 but in one place only. In this way we could explain the mode structure of W1 but the overall observation of Fig. 1 points clearly out two kinds of phase-space features remaining unexplained: i) The mode 3 influences on the unstable manifold of W2 that, at a higher $\mu_C$ value, will manifest also on the W2 orbit itself and which denote a mixing influence from the modes associated with the node towards those associated with the saddle. ii) The presence of all the oscillation modes and their mixing before the creation of the corresponding periodic orbits with their respective invariant manifolds. Thus, the idea that each oscillation mode emerges with the corresponding periodic orbit and that its oscillations extend their influence along the unstable manifold of this orbit is not enough for explaining the oscillatory scenario and either additional or, perhaps, more general mechanisms are needed. The same view will be confirmed and generalized to a larger number of oscillation modes by the subsequent examples with $N = 6$ systems.

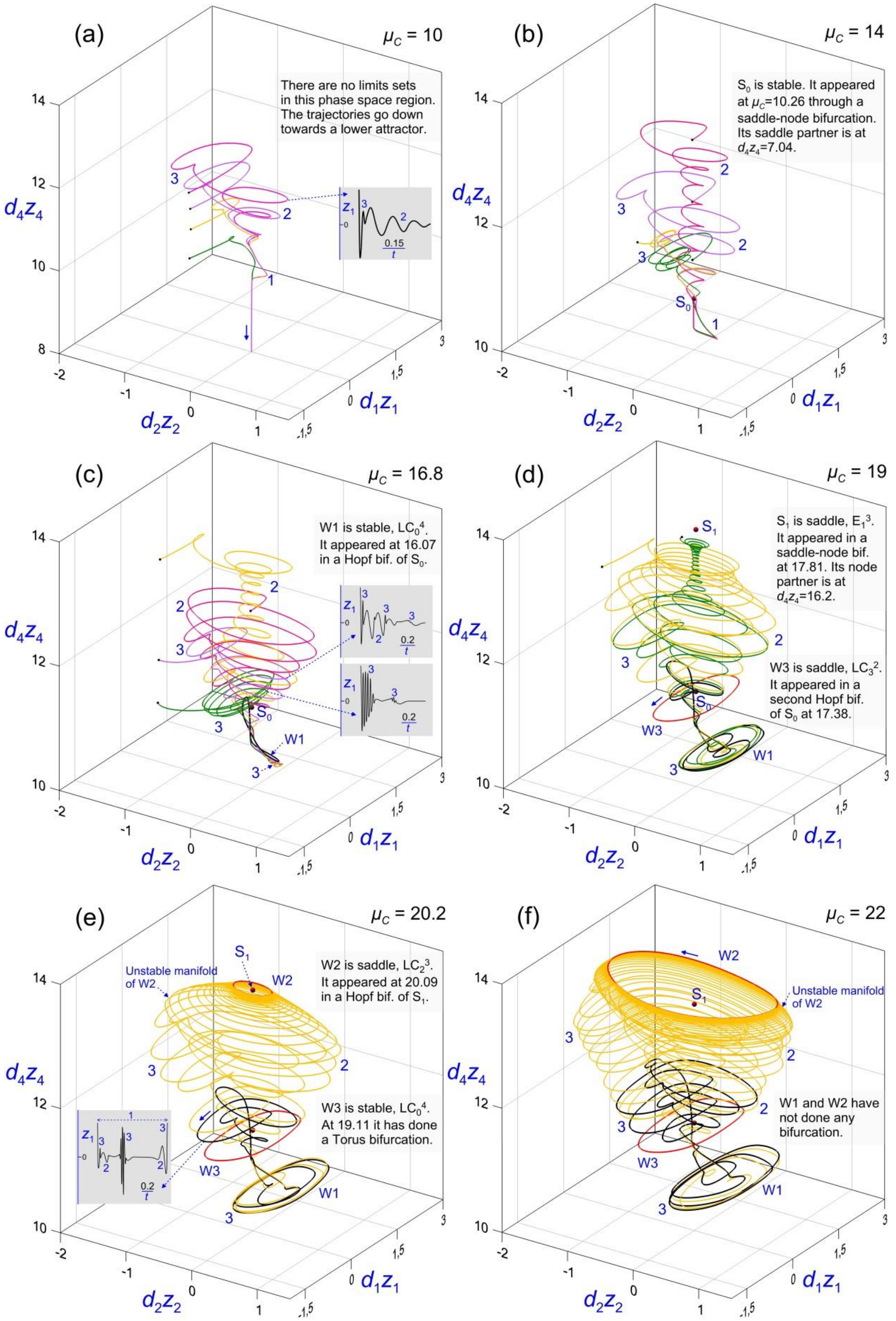

**FIG. 1.** Mixing of three oscillation modes (labeled by numbers) shown in three-dimensional projections of several trajectories for different $\mu_C$ values, for the system $N = 4$ with $g_A$. Black dots denote transient initial points. $S_0$ and $S_1$ denote fixed points, and W$j$ a periodic orbit born with frequency $\omega_j$ from a fixed point. $E_j^q$ and $LC_j^q$ denote equilibria and limit cycles with unstable and stable manifolds of dimension $j$ and $q$, respectively. The first periodic orbit born from $S_0$ is stable (W1, in black) while the others are born as saddles (in red). W1 is influenced by W2 and W3. W2 will be influenced by W3 at higher $\mu_C$ values. The insets show the time evolution $z_1(t)$ of certain trajectories (when comparing notice that $d_1 < 0$).

The analysis of phase portraits is facilitated by considering the following details:

- The fixed points appear located on the $z_N$ axis and, according to Eq. (A5), their coordinate $d_N z_N$ as a function of $\mu_C$ depends exclusively on $g(\psi)$. Figure A1 shows the stationary solutions for the two nonlinear functions, $g_A$ and $g_B$, and indicates the specific fixed points, $S_0$ and $S_1$, involved in the scenario development.
- Every system has been designed such that, when ordered according to their frequency from lower to higher, the various Hopf bifurcations will alternatively occur on $S_0$ and $S_1$ and, on the other hand, the successive bifurcations of a given fixed point will also occur ordered from lower to higher frequency. This means that, if W*j* denotes the periodic orbit born at a Hopf bifurcation of frequency $\omega_j$, the W*j* with *j* odd will arise from $S_0$ and those with *j* even will arise from $S_1$ and, on the other hand, the W*j* emerged from a given fixed point will appear ordered according to *j*. The design procedure generically assures the initial stability of $S_0$ and then $S_1$ appears as a saddle with a one-dimensional unstable manifold. Thus, if $LC_j^q$ denotes a limit cycle with unstable and stable manifolds of dimension *j* and *q*, respectively, and if the Hopf bifurcations are supercritical, as it is the case for all the bifurcations of the three studied systems, the lowest-frequency orbit W1 will appear from $S_0$ as $LC_0^N$ and the rest of orbits W*j* will appear, either from $S_1$ or $S_0$, as $LC_j^{N\text{-}j+1}$.
- The frequency values, and respective periods, imposed to design the system families are given in Table A1.

- The variables $z_j$ have different frequency sensitivity owing to their sequential differentiation relation in the standard form. Such a relation implies that the relative presence of the various oscillation modes in the time evolutions $z_j(t)$, $j = 1, 2, .., N$, increases in proportion to their frequencies each time the subscript $j$ is decreased in one. Thus, $z_1$ optimizes the observation of faster frequencies while they will be practically imperceptible in $z_N$. This fact makes the axes choice strongly influent in what modes appear more pronounced in the phase space projections.
- The various limit cycles emerge from the fixed points within planes whose orientation is exclusively determined by the oscillation frequency of the corresponding Hopf bifurcation. As expressed by Eq. (A9), such a diversity of orientations reflects the different frequency sensitivity of the standard system variables. With so clearly different frequencies as those of the studied systems, the various periodic orbits should appear clearly distinguished by their orientations in the $N$-dimensional phase space, although such a differentiation significantly weakens in the projections used to visualize the phase portraits.

As it is well known, the coexistence of so different timescales implies numerical problems. We used the Runge-Kutta-Fehlberg method with algebraic order of seven and eight and with step-size control. The numerical error influences on the integration of trajectories are particularly manifested in the stable orbit W1 through the loss of strict periodicity without occurrence of any bifurcation. This orbit incorporates the full variety of oscillation modes in high abundance and its periodicity losses begin by slightly affecting the fastest-frequency oscillatory burstings of $z_1(t)$, usually in a $\mu_C$-range where the orbit continuation is even working and the absence of bifurcations may be verified. The location of periodic orbits, usually done through the shooting method although the Poincaré map method worked occasionally better, also manifests troubles due to the multiple timescales. In general, the more influence of faster modes on a given periodic orbit, the more difficult its location and continuous following become. Thus, the location of W1 is the hardest while that of W($N$-1) is the easiest and always easily done. The location troubles increase with the dimension $N$. In our trials up to $N = 6$, all the W$j$ have been located near their birth and then continuously followed along a more or less extended $\mu_C$-range, but for $N = 7$ our software seems unable to operate well in locating W2 and W3, while it is even able with the periodic orbits of faster frequency. On the other hand, concerning the invariant manifolds of periodic orbits, our numerical tools only permit the calculation of those submanifolds associated with either a real multiplier or a complex conjugate pair of multipliers.

The bifurcation diagram of Fig. 2 corresponds to the family of systems with $N = 6$ and the nonlinear function $g_A$. The first thing to be noticed is that the Hopf bifurcations effectively occur at the $\mu_C$ values where the fixed points acquire the $p$ values imposed in the system design and with the frequencies equal to the respectively chosen $\omega_j$ values, and this happens at the two sides of the stationary branches so that the bifurcations occur by pairs with one representing the reverse of the other. All the Hopf bifurcations are supercritical and the numerically-computed characteristic multipliers of the emerging periodic orbits, W$j$, $j$ = 1, 2, .., $N$-1, generically behave as follows when varying $\mu_C$ : i) One multiplier equal to 1 that remains unaltered. ii) One multiplier lower than but practically equal to 1 just after the bifurcation and which gradually decreases when going far from it. iii) ($N$-1-$j$) multipliers practically equal to 0 that remain always near to 0. iv) ($j$-1) multipliers of modulus higher than 1 characterizing (for $j > 1$) the $j$-dimensional unstable manifold of the emerging orbit and which will sustain its torus bifurcations. Thus, except W($N$-1), the rest of periodic orbits have the peculiarity of containing a highly stable component in their stable manifolds. The second thing to be noticed in the diagram is the abundance of torus bifurcations on some periodic orbits and the peculiar fact that their frequencies are practically equal to those of the Hopf bifurcations of the fixed point from which the bifurcating orbit has emerged. For instance, the four torus bifurcations of W5 occur with a secondary frequency $\omega_{sec}$ = 2.071, 0.06284, 0.06284, 2.061, respectively, as given by $\omega_{sec} = sin^{-1}(Im\lambda)/T_{orbit}$, where λ denotes the pair of characteristic multipliers achieving $|\lambda| = 1$ and $T_{orbit}$ is the actual period of the bifurcating orbit. This feature reflects that the considered $\mu_C$ family of systems crosses the space of the dynamical systems relatively near to systems with a pair of saddle-node fixed points both under the eigenvalue degeneracy denoting the simultaneous occurrence of all their Hopf bifurcations, i.e., two on the saddle and three on the node point. It is known that such a kind of degeneracy is the origin of torus bifurcations on the limit cycles appeared from some of the Hopf bifurcations and with secondary frequencies practically equal to those of other Hopf bifurcations of the same fixed point [14]. Although the invariant tori have not been located it is clearly expectable that the associated two-frequency limit sets with their invariant manifolds should participate in the nonlinear mode mixing. Notice that such a kind of torus bifurcation does not introduce new oscillatory modes but only additional mixing mechanisms. Notice also that, as happens with the Hopf bifurcations of a fixed point, each torus bifurcation of a periodic orbit is accompanied by the reverse one at a higher $\mu_C$ value. Our numerical trials with systems in the form of Eqs. (1) and (2) indicate that these torus bifurcations occur at lower frequency on limit cycles of faster frequency, while the faster frequencies affect the limit cycles of slower frequency through the mechanisms of nonlinear mode mixing. Notice that unlike the mode mixing, which directly modifies the limit cycles and their oscillations, the torus bifurcations only alter their stability while creating the new two-frequency limit sets. It is also worth remarking that, although relatively generic, the torus bifurcations do not always occur. See, for instance, the bifurcation diagram of Fig. 8 corresponding to the same $N$=6 system but with another nonlinear function, in which the torus bifurcations are absent. In the diagram of Fig. 2 all the torus bifurcations happen within the unstable manifold of the bifurcating orbit so that it increases its stability in two dimensions at each bifurcation and, in some cases, the orbit can become stable. Notice the coexistence in a certain $\mu_C$ range of up to three attractors: W5, W3, and that derived from W1, being the various attraction basins supposedly delimitated by the five-dimensional stable invariant manifolds of the limit sets born with the corresponding invariant tori.

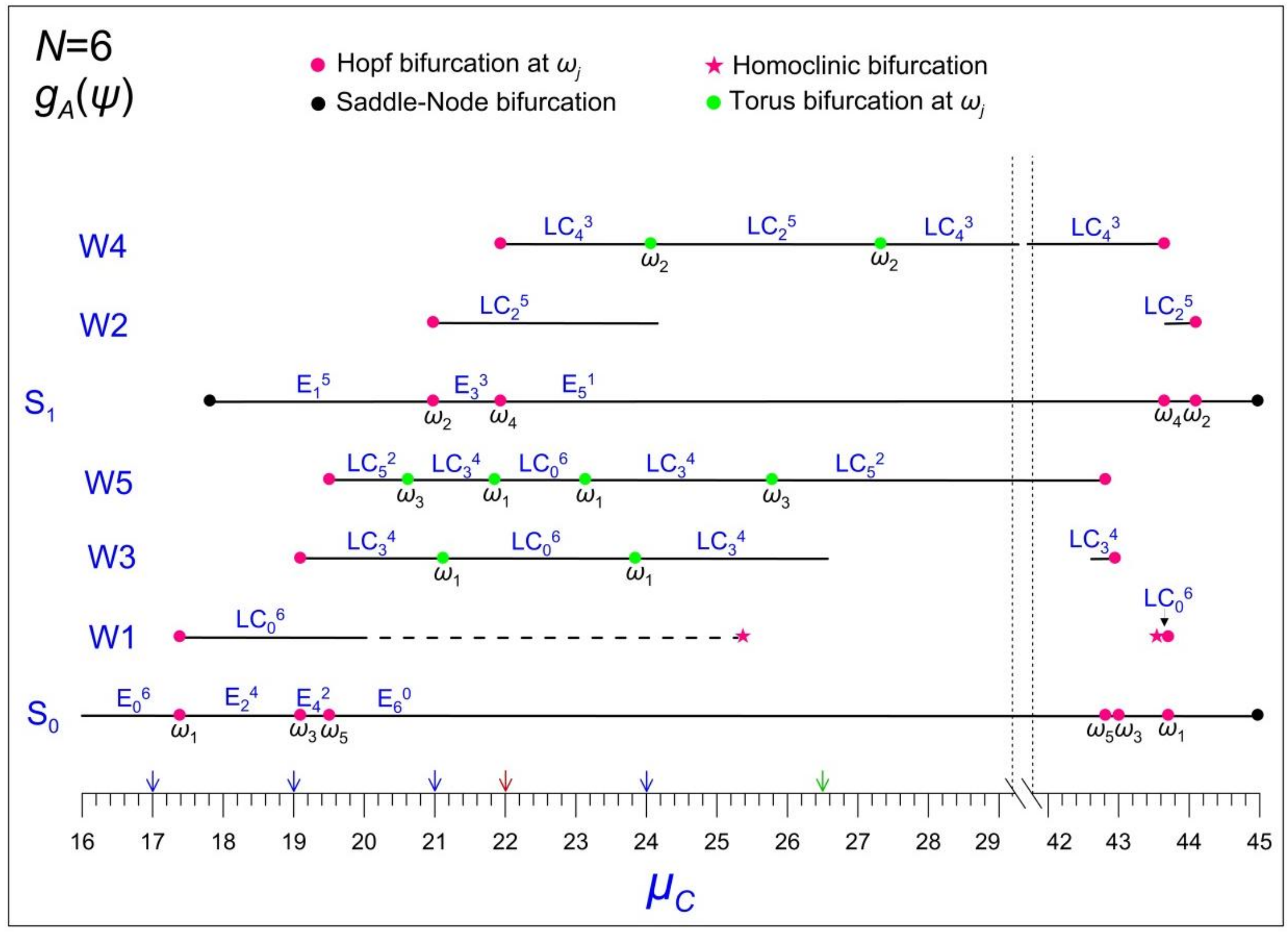

**FIG. 2.** Bifurcation diagram of the $\mu_C$-family of systems with $N = 6$ (Table A1) and the nonlinear function $g_A$ (Eq. (A7)), concerning the saddle-node pair of fixed points $S_0$ and $S_1$ of Fig. A1 and the periodic orbits W$j$ born from these points. $E_j^q$ and $LC_j^q$ denote equilibria and limit cycles with unstable and stable manifolds of dimension $j$ and $q$, respectively, as determined from the set of eigenvalues of each limit set. The continuous line denotes location of the corresponding limit set and the broken line in W1 indicates an attractor derived from the initial periodic orbit. Quasiperiodic orbits have not been located. The blue, red and green arrows denote the $\mu_C$ values of the phase portraits shown in Figs. 4, 5 and 7, respectively.

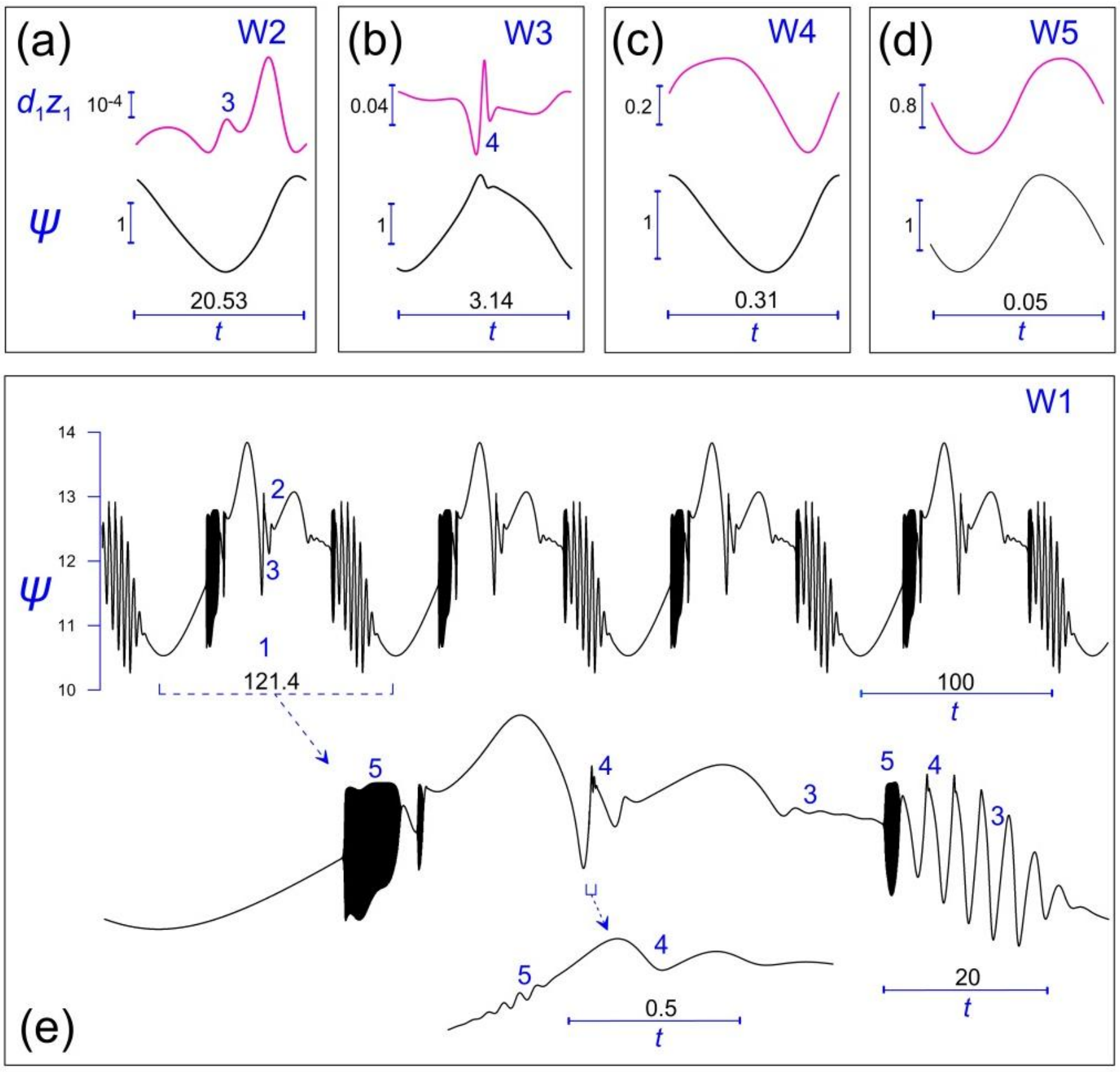

**FIG. 3.** Time evolutions associated with the five periodic orbits emerged from a saddle-node pair of fixed points in the $N = 6$ system with $g_A$ , for $\mu_C = 23$. Notice the nonlinear mode-mixing influence on the various orbits, especially remarkable in W1 but also clear on W2 and W3. Generically, the mixing occurs through the incorporation of faster modes on slower ones and the period of the influenced orbit usually increases with respect to that of the Hopf bifurcation. From the diagram of Fig. 2 it is seen that W3 and W5 are also stable at this $\mu_C$ value, in addition to W1. W1 looks practically periodic, but in phase space representations involving the faster variables it is seen that the successive cycles do not superpose well the $\omega_5$ oscillations.

As it has been said, the location and continuation of W1 is the most critical. Sometimes its location has been achieved after computing a long asymptotic transient towards it and then choosing a proper point far enough from any fast oscillation for initiating the locating algorithm. The continuous line in the bifurcation diagram of Fig. 2 denotes where we are confident that the periodic orbit remains without having doing any bifurcation while the broken line denotes the asymptotic attractor derived from W1 when the periodicity loses become more pronounced. Nevertheless, such loses decrease again and at $\mu_C = 23$ the asymptotic signal becomes practically periodic, as can be seen in Fig. 3. At $\mu_C = 23.75$ the signal points clearly out the occurrence of a period doubling and the consequent development of chaos up to the attractor destruction in a homoclinic bifurcation.

The behavior of this $\mu_C$ family of systems is illustrated in the three-dimensional projections of phase portraits shown in Figs. 4 and 5. Like in the $N = 4$ case, the variety of trajectories clearly illustrate how the several oscillation modes and their mixing manifest over extended zones of the phase space and how this happens even before the appearance of the fixed points from which the corresponding periodic orbits will emerge, and also how the mode mixing affects the periodic orbits themselves. A variety of initial points have been chosen to provide a general overview of the transient main features but a more systematic analysis is of course pending. Concerning the periodic orbits notice in particular how a burst of $\omega_4$ oscillations appear and then disappear on W2 (portraits at $\mu_C = 21$, 22 and 24) suggesting that the growing periodic orbit incorporates the $\omega_4$ oscillations when it crosses a region where such oscillations occur. More details about the transformation of W2 as a function of $\mu_C$ are given in Fig. 6 where the orbit shows the incorporation of oscillations of the three faster frequencies. The mode-mixing influence on the periodic orbits is a rather general feature of the oscillatory scenario, as is shown in Fig. 7 for a higher $\mu_C$ value, at which the orbits W3 and W4 have also incorporated faster frequency oscillations. Thus, in general the periodic orbits incorporate intermittent influences of other oscillation modes by maintaining their self-sustained attribute and without experiencing any bifurcation. Typically, a periodic orbit of a given frequency develops a gradual incorporation of all the faster modes, associated either with the same fixed point from which it has appeared or with the saddle-node partner, and particularly remarkable is the twofold sense of the mixing influence from the modes of the saddle towards those of the node and vice versa. In particular, the lowest-frequency orbit W1 incorporates influences of all the other modes in high abundance and with a rather complex combination of ones within the others.

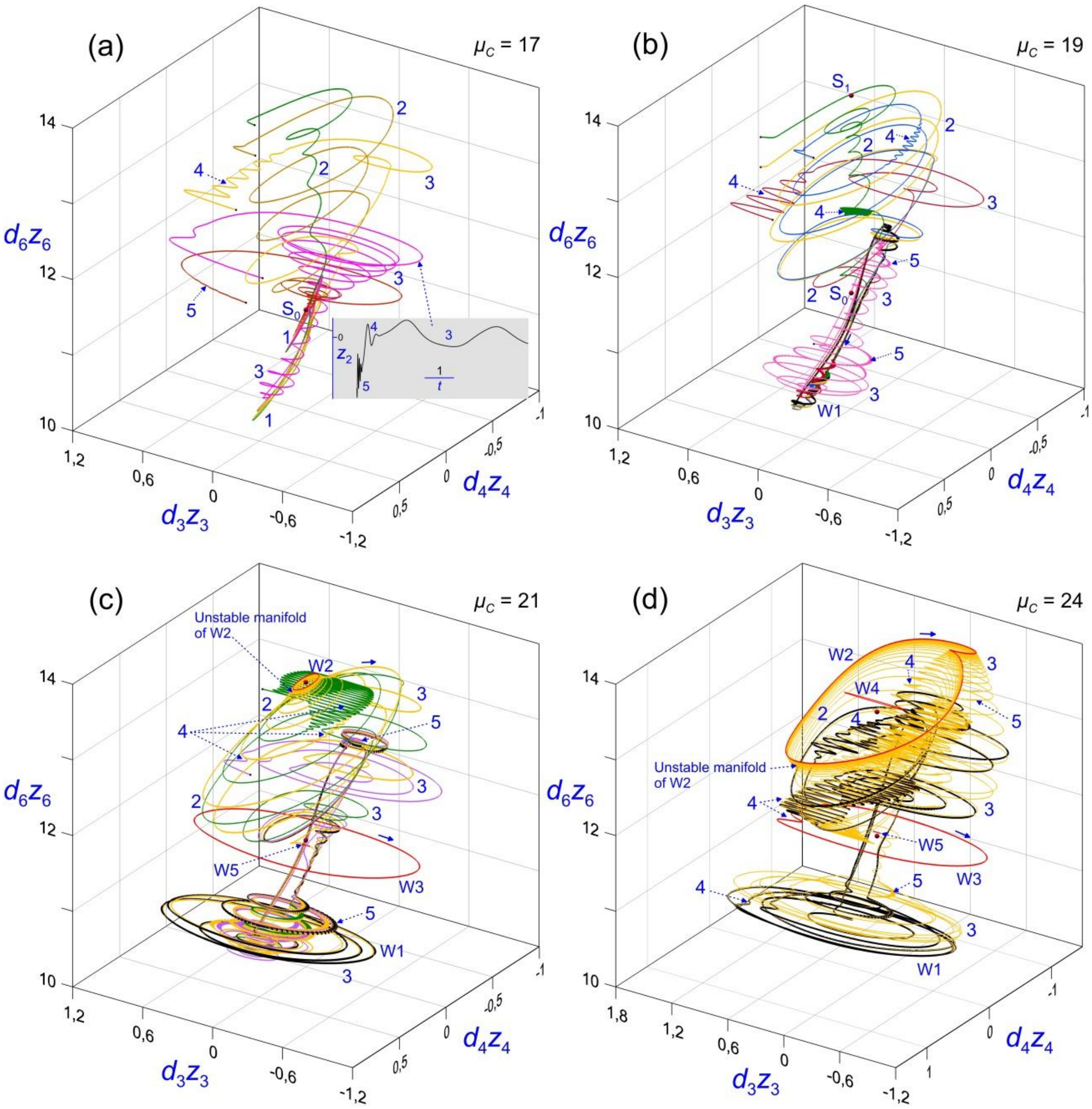

**FIG. 4.** Nonlinear mixing of five oscillation modes in the $\mu_C$-family of $N$=6 systems with the nonlinear function $g_A$. The bifurcations of fixed points and periodic orbits are indicated in the diagram of Fig. 2. The two-dimensional unstable manifold of W2 is represented through a single trajectory. Twenty such trajectories initiated from different points along the periodic orbit have been computed; all of them look rather similar and they simply fill the two-dimensional surface by maintaining the structure that is better appreciated with a single trajectory. The non-represented coordinates of the transient initial points may have non-zero values. At $\mu_C$ = 21 and 24, the orbit W5 is $LC_3^4$, and it should be surrounded by the $(\omega_5, \omega_3)$ two-frequency limit set born at the corresponding torus bifurcation and whose unstable manifold would be also involved in the mode mixing. At $\mu_C$ = 24, the attractor W1 is chaotic but only one cycle has been represented for clarity.

Finally, we deal with the $\mu_C$ family of $N$ = 6 systems with the same set of $c_j$ and $d_j$ coefficients as before (see Table A1) but with a Gaussian nonlinear function, Eq. (A8), instead of the periodic interferometric function, Eq. (A7). The corresponding diagram of bifurcations shown in Fig. 8 points out that this $\mu_C$ family does not cross any torus bifurcation, while the projected phase portraits of Fig. 9 illustrate how the mode mixing works also in this family.

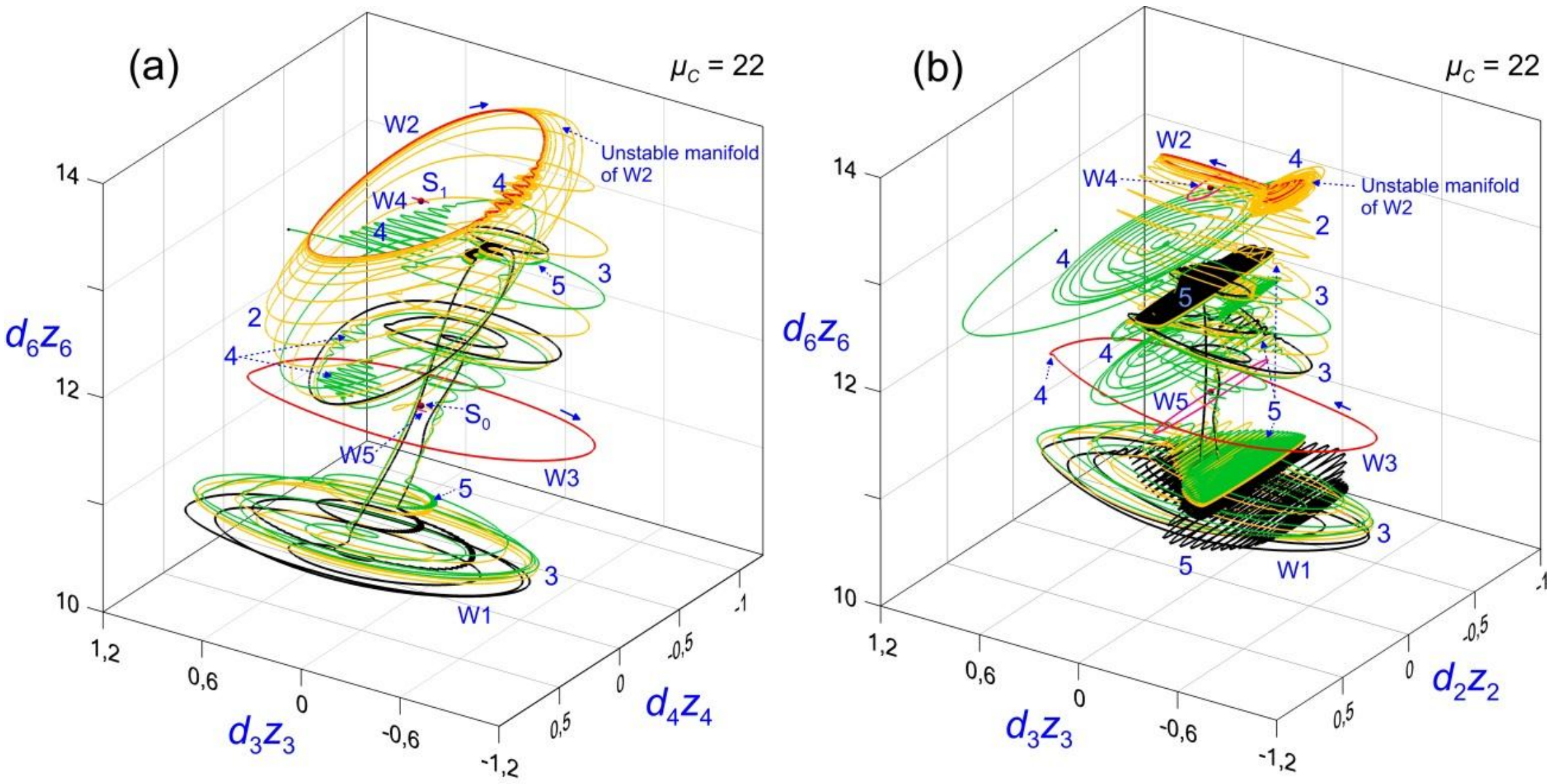

**FIG. 5.** The same as in Fig. 4 for a different $\mu_C$ value, with the representation of (a) corresponding to the same three-dimensional subspace while the projection of (b) is on the coordinate $z_2$ instead of $z_4$. It illustrates how the different frequency sensibility of the variables influences the phase-portrait visualization. Concretely, with respect to $z_4$, $z_2$ improves the relative presence of the oscillations at $\omega_4$ and $\omega_5$, leaves unaffected those at $\omega_3$, and decreases those at $\omega_2$ and $\omega_1$. In fact, concerning $z_2$, the several trajectories lack of noticeable amplitude modulation at $\omega_1$ and $\omega_2$. At this $\mu_C$ value, both W3 and W5 are stable, with the former probably surrounded by a ($\omega_3$, $\omega_1$) quasiperiodic orbit and the latter probably surrounded by two quasiperiodic orbits of frequencies ($\omega_5$, $\omega_1$) and ($\omega_5$, $\omega_3$), respectively.

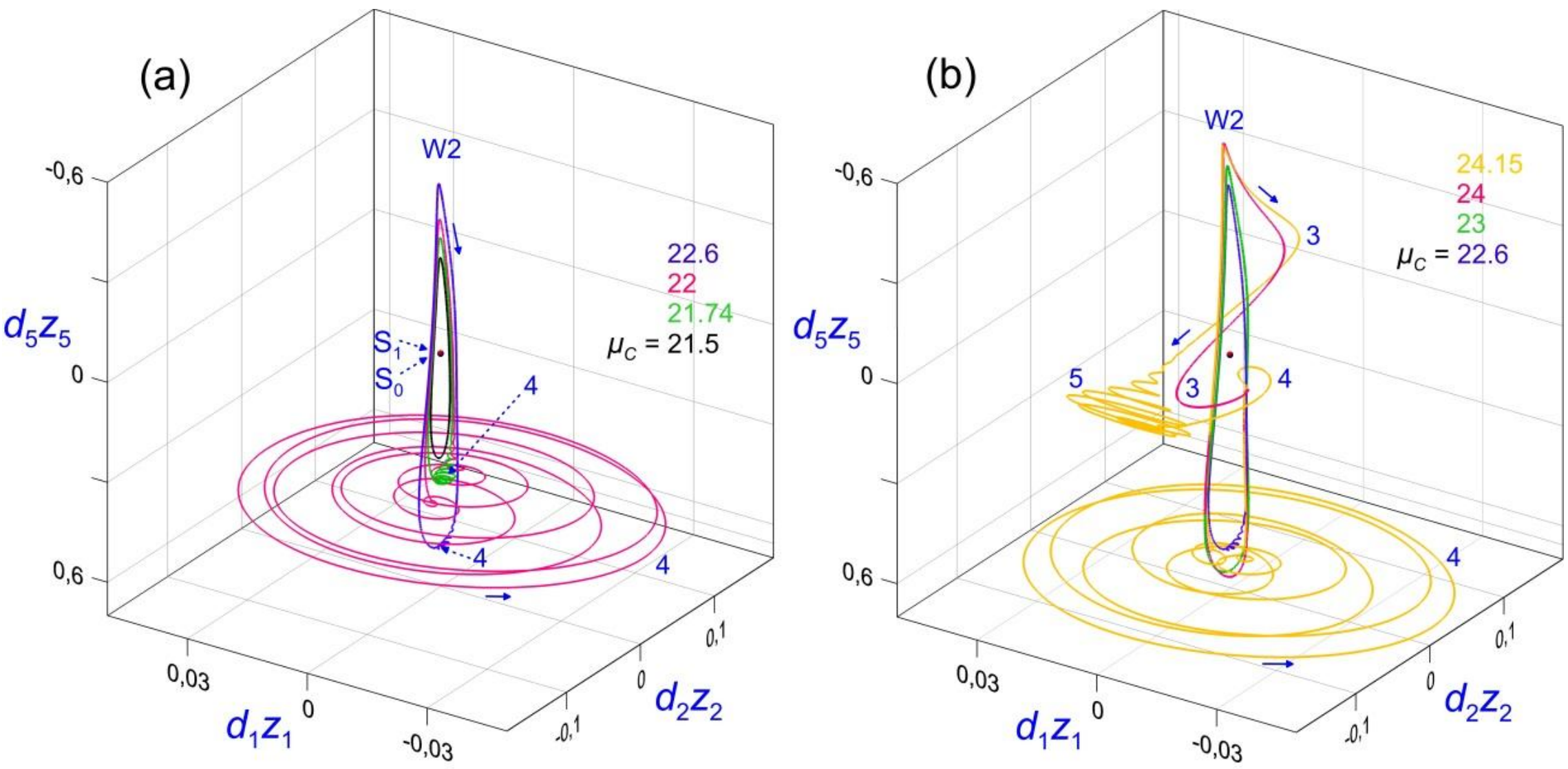

**FIG. 6.** Periodic orbit W2 of the $N = 6$ system with $g_A$ for different $\mu_C$ values. Unlike in previous representations now the vertical axis is not $z_N$ so that both fixed points appear located on the origin independently of the $\mu_C$ value. Notice how the orbit transforms by incorporating localized oscillation bursts of modes 4, 3, and 5, and all of these changes happen to the periodic orbit without doing any bifurcation (see the bifurcation diagram of Fig. 2). Notice also how the presence of mode 4 first increases, then decreases, and then increases again as a function of $\mu_C$.

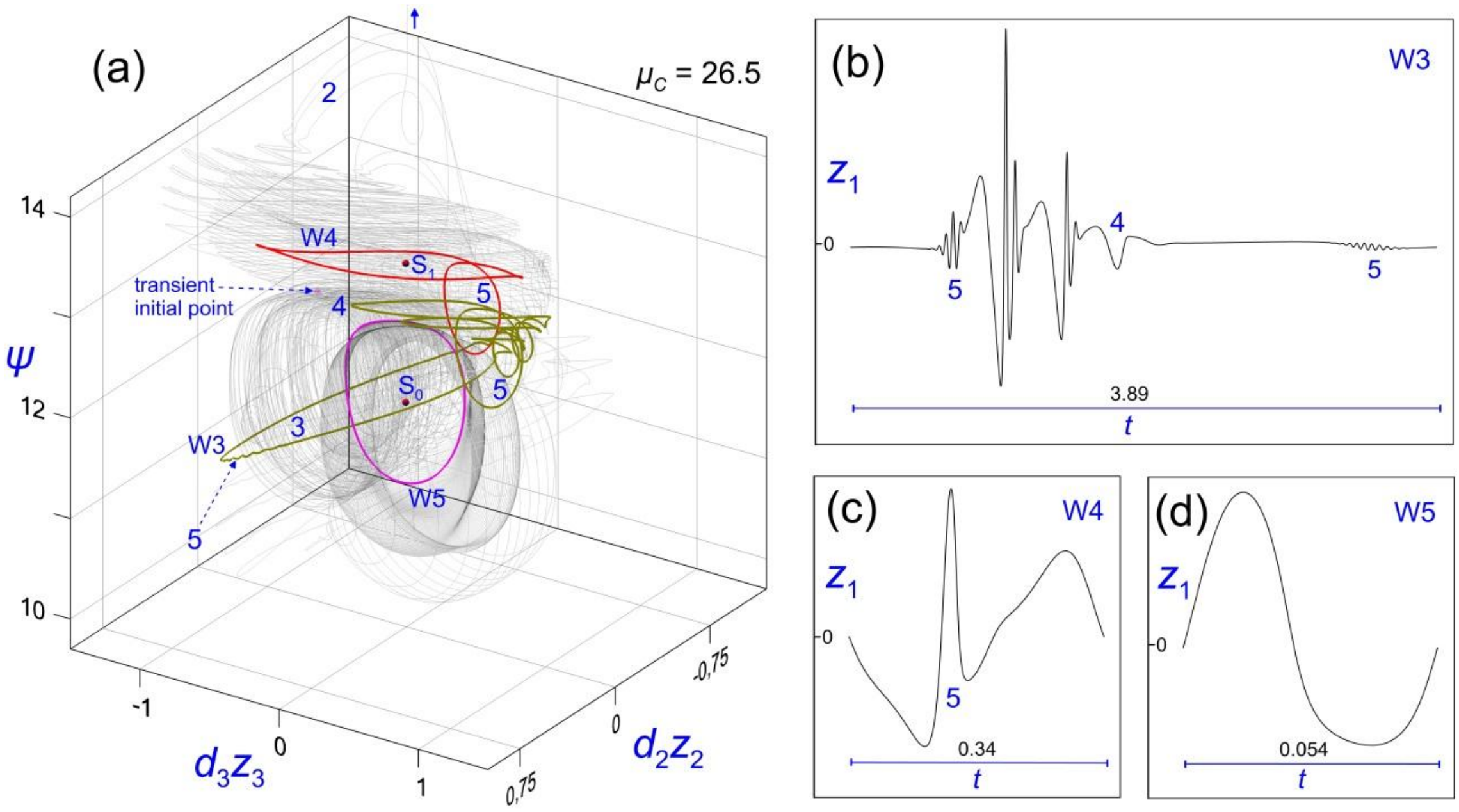

**FIG. 7.** Illustration of how the nonlinear mode mixing also affects the high-frequency periodic orbits, for the $N = 6$ system with $g_A$. The representations correspond to the maximum $\mu_C$ value of the W3 continuation, at which the attractor derived from W1 has disappeared and W2 has not been located, although it is expected to exist. (a) Phase-portrait projection in which the vertical scale $\psi$ is sensitive to all the oscillation frequencies and then reinforces the higher frequency modes. W3 and W5 are $LC_3^4$ and $LC_5^2$, respectively, while W4 is $LC_2^5$, and it should be surrounded by a quasiperiodic orbit $QP_4^4$. The transient in black line illustrates how the oscillatory mixing affects the phase space even in the absence of any attractor. (b -d) Time evolutions $z_1(t)$ of the three periodic orbits along one period illustrating the mode mixing at the timescale. Notice the relative absence of amplitude modulation at $\omega_3$ in the $z_1(t)$ *of* W3.

## III. Concluding comments

Three things are worth to be remarked on in the light of the reported portraits and corresponding time evolutions. First, the various oscillatory modes appear in extended regions, either on the transient trajectories or on the periodic orbits, by maintaining their frequency and their phase-space orientation everywhere. This means that each mode describes a well-defined dynamical activity among the various variables since the orbit orientation defines their relative participation according to the orbit projection on the respective axes. In words, the various modes describe a variety of specific and characteristic dynamical activities and, therefore, their mixing describes a peculiar combination of such specific activities for each one of the large variety of phase-space trajectories. We expect that such a relevant feature would be a generic one of the generalized Landau scenario, i.e., even when developed in dynamical systems of arbitrary kind, while the observed fact in our examples that both the mode frequency and the orientation remain also almost equal in all the systems of a given $\mu_C$ family should be seen as a peculiar feature of the considered kind of systems, Eqs. (1).

Second, the intermittent incorporation of faster oscillations within a slower one often increases the period of the latter. This is particularly manifested in the periodic orbits but also happens in the transients when crossing regions of strong mode mixing.

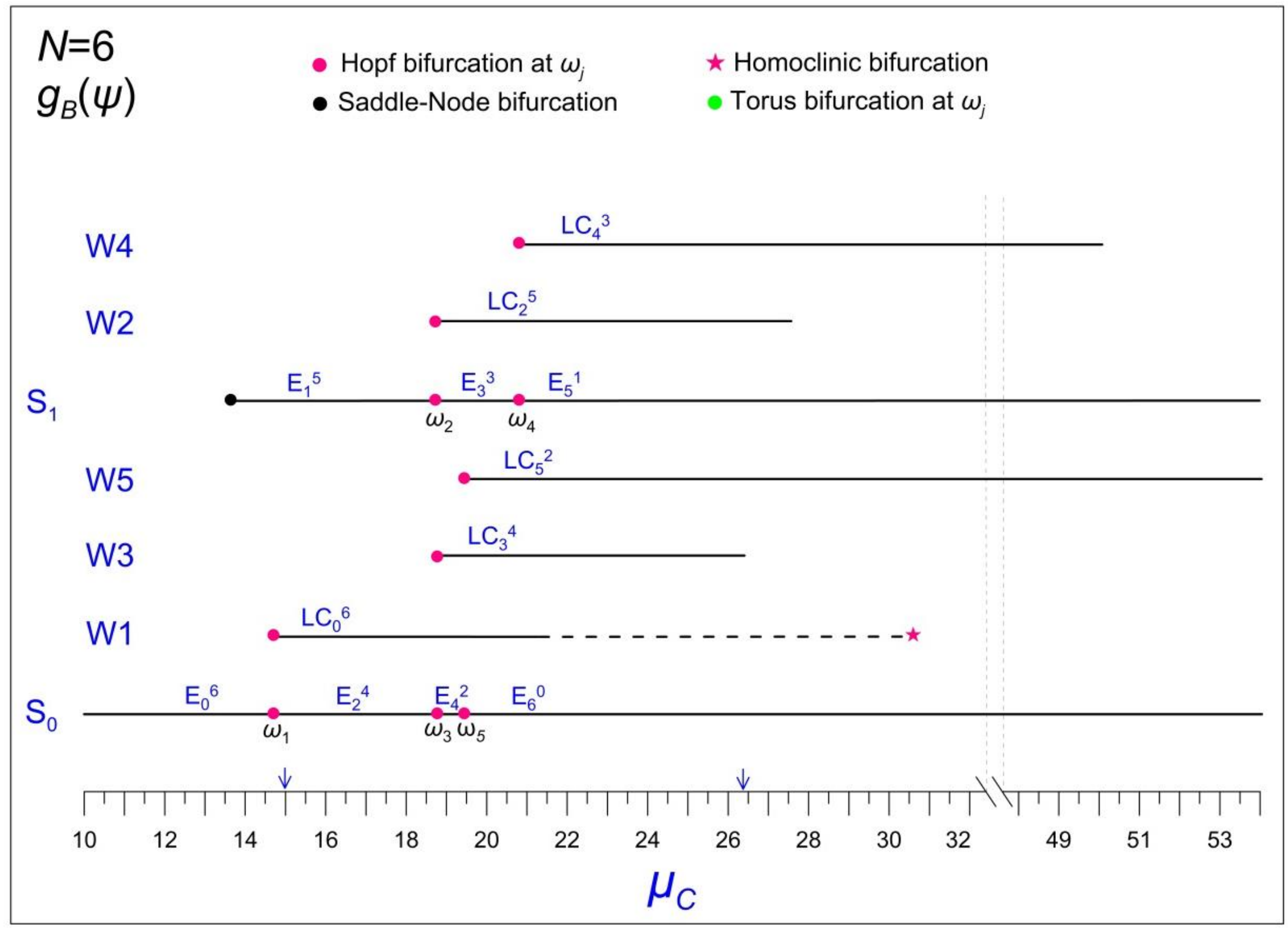

**FIG. 8.** Bifurcation diagram for the $\mu_C$ family of $N = 6$ systems with the same $c_j$ and $d_j$ coefficients as before but with another nonlinear function, $g_B$ [Eq. (A8)]. The diagram does not cover the right-hand side of the stationary diagram since it extends up to near $\mu_C = 100$. The main difference with respect to the previous case is the absence of torus bifurcations. In fact, except W1 in its approach to the homoclinic bifurcation, the other periodic orbits do not suffer any bifurcation along the respective $\mu_C$ ranges of their continuous following. Another difference is in the relative position on the $\mu_C$ scale of the Hopf bifurcations of the two fixed points, with those of $S_1$ more superposed here on those of $S_0$.

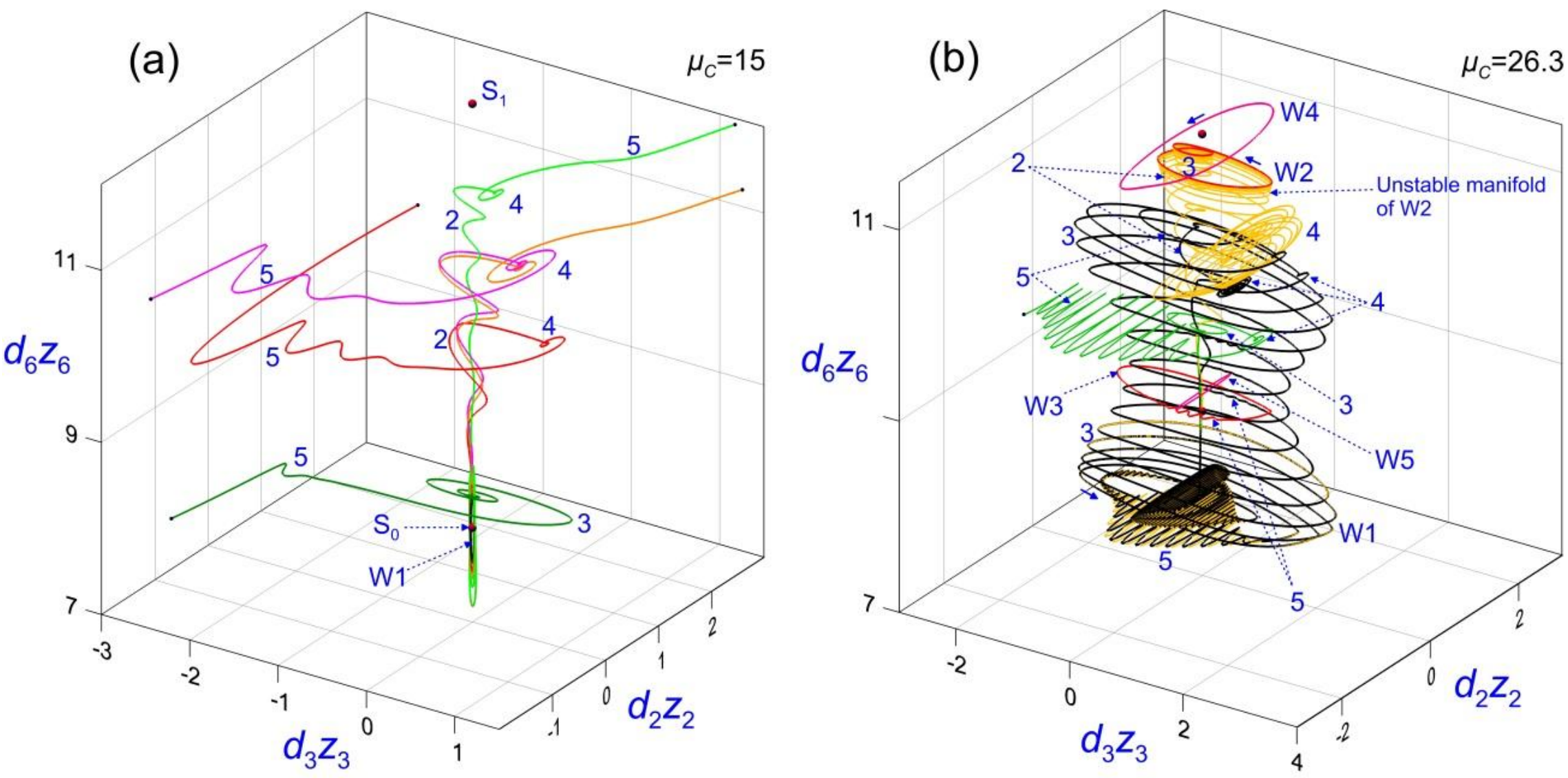

**FIG. 9.** Transients and limit sets in three-dimensional projections of phase portraits for the $\mu_C$ family of $N = 6$ systems with $g_B$, for two values of $\mu_C$. The most peculiar feature with respect to before is the abundance of $\omega_3$ oscillations in W1.

It should be stressed however that the period enlargement does not always occur. Concretely, it does not happen in the mixing among modes emerged from the same fixed point [12], and this may be related to the fact that, in this circumstance, the mixing does not involve any homoclinic process. Instead the period enlargement generically occurs when the mixed modes have emerged from the saddle-node pair of fixed points, either when the influencing faster mode appears from the saddle or when it appears from the node point. In these circumstances the involved periodic orbits usually admit to be compatible with participating in some homoclinic process [12]. In any case, the nonlinear mode mixing onto a periodic orbit happens while the orbit remains periodic, i.e., single periodic, and, this should be confronted with the multiperiodic limit sets arising from the torus bifurcations. For instance, in either the two-frequency periodic orbits or the quasiperiodic orbit arising from a two-torus bifurcation, the secondary oscillation is continuously (and initially linearly) combined with the primary oscillation along the entire bifurcating periodic orbit and this is both the source of the so-called resonance problems of the torus bifurcation and the reason of why the bifurcation occurs with all of its details in the discrete system defined by a Poincaré section of the continuous one. In contrast, it is worth noting the absence of any resonance problem in the mode-mixing mechanisms, because they nonlinearly incorporate the secondary oscillation in localized zones of the primary orbit only, as well as the incapability of any Poincaré section to capture the intermittent mode mixing in the single-periodic orbit, because it maintains its stability untouched and no additional invariant sets are created. The absence of resonance problems permits the succession of localized incorporations of additional modes and of the same mode several times on different places, while the periodic orbit remains single periodic but with the period becoming usually enlarged and, most importantly, with the associated periodic activity becoming enriched through the succession of dynamical activities of the different oscillation modes, each one of them affecting the system variables in a peculiar way. Of course such a kind of dynamical enrichment happens on all the phase-space trajectories crossing the mode-mixing region, but we consider its occurrence on the periodic orbits as paradigmatically characterizing the essence of the oscillatory scenario development and then we propose the name of *nonlinear complexification* of a periodic orbit to describe the becoming complex of its associated dynamical activity by nonlinear mode mixing, without necessity of any bifurcation and starting from the harmonic oscillation born in the corresponding Hopf bifurcation of a fixed point.

Third, each one of the various oscillation modes affects extended phase space regions, appearing as covering different places in the three-dimensional projections but perhaps connected in the full $N$-dimensional space. Notice in particular the diversity of places where the faster modes appear and how far from the zone of periodic orbits they arrive. The phase space characterization by numerical means would require systematic analyses of transients initiated from some ($N$-1)-dimensional surface and this looks extremely hard. Our limited analysis suggests rather intricate features in the extension of the various oscillation modes for the phase space and in their mixing. The phenomenon is clearly of global nature since it develops independently of the presence of limit sets and other invariant sets but it is also clear that a more attainable analysis can be done by considering the periodic orbits emerged from the saddle-node pair of fixed points and the invariant manifolds of these orbits connecting ones with others. In fact, as already noticed in relation to Fig. 1, our description trying to explain the mixing mechanisms [12,13] is based on how some periodic orbits extend their oscillations along their unstable manifold towards other periodic orbits like a kind of corkscrew effect, through which the dynamical effects associated with the influencing orbit are intermittently incorporated within those of the influenced orbit without altering its stability and self-sustaining balance. This view associating the mode mixing with the unstable manifold of the influencing orbit applies well in a variety of cases but it is not general enough to cover all the circumstances and, of course, it does not apply when the limit sets are lacking. Particularly difficult to be explained in this way is the influence of periodic orbits appeared from the node point towards those emerged from the saddle since the influencing orbits are born with unstable manifolds lacking any connection towards the influenced orbits and its formation during the scenario development seems unlikely. For instance, in the phase portrait of Fig. 4(d) the influence of mode 3 on W2 happens while the three-dimensional unstable manifold of W3 (not drawn in the figure) is fully going towards W1. On its hand, the two-dimensional unstable manifold of W2 descends towards W1 by incorporating oscillations of mode 3 (as well as of modes 4 and 5) upon its own $\omega_2$ oscillations, apparently suggesting that the influence of mode 3 is going from W2 to W3. The generic nature of this kind of mode mixing is pointed out in Figs. 6 and 7 by the presence of a burst of mode 5 on the periodic orbits W2 and W4, respectively. On the other hand, the influences of high-frequency modes on transient trajectories far enough from the attractor cannot be also associated with any unstable manifold of the periodic orbits.

A mathematical theory of the oscillatory mixing scenario should specify the required conditions for the scenario development and characterize the mechanisms sustaining such a development, in which substantial topological aspects would surely be implicated in addition to the bifurcational ones. In essence, such a kind of theory, if successful, would describe how a dynamical system with a large number of degrees of freedom can combine their activities to sustain complex but coordinated behaviors of the whole. According to our experience, we expect that the possession of a saddle-node pair of fixed points experiencing successive two-dimensional instabilities will be enough for the full scenario development in its simplest form.

The final comment must concern the potential relevance of the generalized Landau scenario and this means to realize how the systems exhibiting it extend for the space of the dynamical systems. Starting from one of our $N$-dimensional $\mu_C$-families, it is easy to appreciate the robustness of the oscillatory behavior by verifying how slightly the behavior varies when the values of the $c_q$ and $d_q$ coefficients are gradually and significantly modified or when the $g(\psi)$ function is changed. The starting family should have been designed by firstly choosing the dimension $N$ and then a set of $N$-1 pairs of ($p_j$, $\omega_j$) values, and, by freely and continuously changing the prechosen values, a vast variety of families of arbitrary dimension would be obtained, always exhibiting the generalized Landau scenario in full. For instance, a numerical simulation for $N = 12$ is reported in Ref. [13]. Our conclusion is that there is no reason for a limit in the dimension growing of such a kind of families exhibiting up to $N$-1 oscillation modes on the basis of a saddle-node pair of fixed points, and we are now trying to achieve designable systems with larger sets of fixed points exploiting all their Hopf bifurcation possibilities [15].

In a broader view of the space of the dynamical systems, and reasonably assuming the possession of fixed points able to do successive Hopf bifurcations as sufficient condition for the scenario development, we expect that the generalized Landau scenario should come forth through successive crossings of the two kinds of codimension-one bifurcation surfaces, the saddle-node and Hopf bifurcations, while the oscillatory mixing should happen without requiring any bifurcation. Concerning the occurrence of successive Hopf bifurcations, it is worth stressing that the next Hopf bifurcation is not more demanding than the previous one, since it should be of codimension-one again, and then the sequence of bifurcations does not require more conditions than the sequence of those of the singular bifurcations. Generally speaking, the scenario should develop, and reversely should dismantle, as a gentle process associated with the gradual intertwinement of trajectories through the mode-mixing mechanisms and with the successive incorporation of other fixed points and new periodic orbits. The regions of oscillatory systems should extend in continuity towards higher dimensions, without any disruption at the crossing of additional saddle-node and Hopf bifurcation surfaces or when crossing the densely accumulated bifurcations of chaos. Only certain global bifurcations of homoclinic nature can destroy the attractor but without altering the oscillatory mixing scenario that then will contain transient trajectories eventually evolving towards another basin of attraction. Thus, we foresee that the generalized Landau scenario provides the world of dynamical systems with extraordinary oscillatory possibilities through which indefinitely large numbers of degrees of freedom can combine their activities to sustain rather complex but ordered dynamical behaviors. On the other hand, it is worth noting that, to the best of our knowledge, no alternative mechanisms achieving equivalently complex behaviors are known in nonlinear dynamics and, most importantly, there are no pieces of evidence for suspecting their existence.

## Acknowledgments

This work has been partially funded by Spanish Ministry of Economy and Competitiveness (MINECO) under Grant No. FIS2014-57460P and the Catalan Government under Grant No. SGR2014-1639. The comments and suggestions by three referees helped to significantly improve the presentation of the paper.

**Appendix. System of equations**

A very general description of the $N$-dimensional dynamical systems is

$$\frac{dx}{dt} = Ax + \sum_{j=1}^{m} b_j f_j(x, \mu), \qquad \text{(A1)}$$

where $x \in \Re^N$ is the vector state, $A$ is a constant $N$x$N$ matrix, $b_j$ are constant $N$-vectors, $f_j$ are scalar-valued functions nonlinear in $x$, $\mu$ describes constant parameters involved in the nonlinear functions, and the $m \leq N$ components $b_j f_j$ are linearly independent. Under appropriate nonlinearities, the system (A1) may possess $m$-dimensional arrays of fixed points and a basin of attraction can involve up to $3^m$-1 saddle fixed points of different types in addition to the attracting one [12] .

For $m = 1$ and provided that the matrix ($b_1$, $Ab_1$, $A^2b_1$,... , $A^{N-1}b_1$) has rank equal to $N$, system (A1) can be linearly transformed in a standard form like

$$\begin{aligned} \frac{dz_1}{dt} &= -\sum_{q=1}^{N} c_q z_q + f_1(z, \mu), \\ \frac{dz_j}{dt} &= z_{j-1},\ j = 2, ..., N, \end{aligned} \qquad \text{(A2)}$$

where $z$ is the new vector state and $z_q$ its components. The fixed points would appear located on the $z_N$ axis. The system design [10] is facilitated by considering nonlinear functions of a single variable in the form

$$f_1(z, \mu) = \mu_C\, g(\psi, \mu), \qquad \text{(A3)}$$

with

$$\psi = \sum_{q=1}^{N} d_q z_q, \qquad \text{(A4)}$$

and where $\mu_C$ will be taken as a control parameter. Without loss of generality we assume $c_N = d_N$ since the transformation $z_q \to (d_N/c_N) z_q$, $d_q \to (c_N/d_N) d_q$, $c_q \to c_q$, and $\mu_C \to (d_N/c_N)\mu_C$ leaves system (A2-A4) invariant but with $d_N = c_N$. The equilibria of system (A2-A4) should fulfil the conditions

$$\begin{aligned} \bar{z}_{q \neq N} &= 0, \\ \bar{\psi} = d_N \bar{z}_N &= \mu_C\, g(\bar{\psi}), \end{aligned} \qquad \text{(A5)}$$

where the overline denotes steady-state values. Notice that the steady-state solution $\bar{\psi}(\mu_C)$ is independent of the dimension $N$ and the coefficients $c_q$ and $d_q$, while it is exclusively determined by $g(\psi)$. On the other hand, it may be shown [10] that the influence of $\mu_C$ and $g(\psi)$ on the linear stability of the equilibria passes exclusively through the value of the parameter

$$p(\bar{\psi}) = \mu_C \left[ \frac{\partial g}{\partial \psi} \right]_{\bar{\psi}}, \tag{A6}$$

which is also independent of $N$, $c_q$, and $d_q$. This means that the linear stability can be handled without specifying a concrete nonlinear function and therefore without knowing the actual fixed points but identifying them by means of the corresponding hypothetical values of $p$. By properly choosing the corresponding set of pairs of values for $p$ and the frequency $\omega$, we can impose the occurrence of up to $N$-1 Hopf instabilities in a hypothetical saddle-node pair of fixed points and, in this way, determine the set of appropriate coefficients $c_q$ and $d_q$ with which the fixed points of system (A2-A4) will experience the various Hopf bifurcations with the chosen frequencies at the $\mu_C$ values where they reach the corresponding $p$ values. Additionally, we want systems possessing attractor and to achieve it the design procedure should be properly constrained [10].

The main requirement on the nonlinear function $g(\psi)$ is that it should describe some sort of dip or hump to allow for the coexistence of a saddle-node pair of fixed points with proper values for their parameter $p$, while its detailed expression would have a secondary, although of course relevant, influence on the oscillatory behavior. The simulations reported in this article have been done with two different functions:

$$g_A(\psi) = \frac{1.25 - 1.06\cos\psi}{1.68 - \cos\psi}, \tag{A7}$$

$$g_B(\psi) = 1.1 - e^{-\left(\frac{\psi - 10}{2.5}\right)^2}, \tag{A8}$$

of which the first is periodic and describes the interferometric Airy function of the family of physical devices through which the oscillatory scenario was discovered [11], while the second is simply an inverted Gaussian. Figure A1 shows a graphical representation of each nonlinear function together with the corresponding steady-state solution $\bar{\psi}(\mu_C)$ and distribution of $p(\bar{\psi})$ values. The saddle-node bifurcations correspond to $p = 1$ and the arising pairs of fixed points, one with $p > 1$ and another with $p < 1$, would correspond to the saddle and the node, respectively, provided that the initial fixed point at $\mu_C = 0$ would be stable.

The reported simulations correspond to the $N = 4$ and $N = 6$ systems described in Table A1. In the design process, different enough frequencies for the several oscillation modes have been chosen in order to facilitate their identification on the time evolution signals. The nonlinear mode mixing works also for more similar frequencies but the waveform structures become then gradually blurred and their analysis is rather difficult.

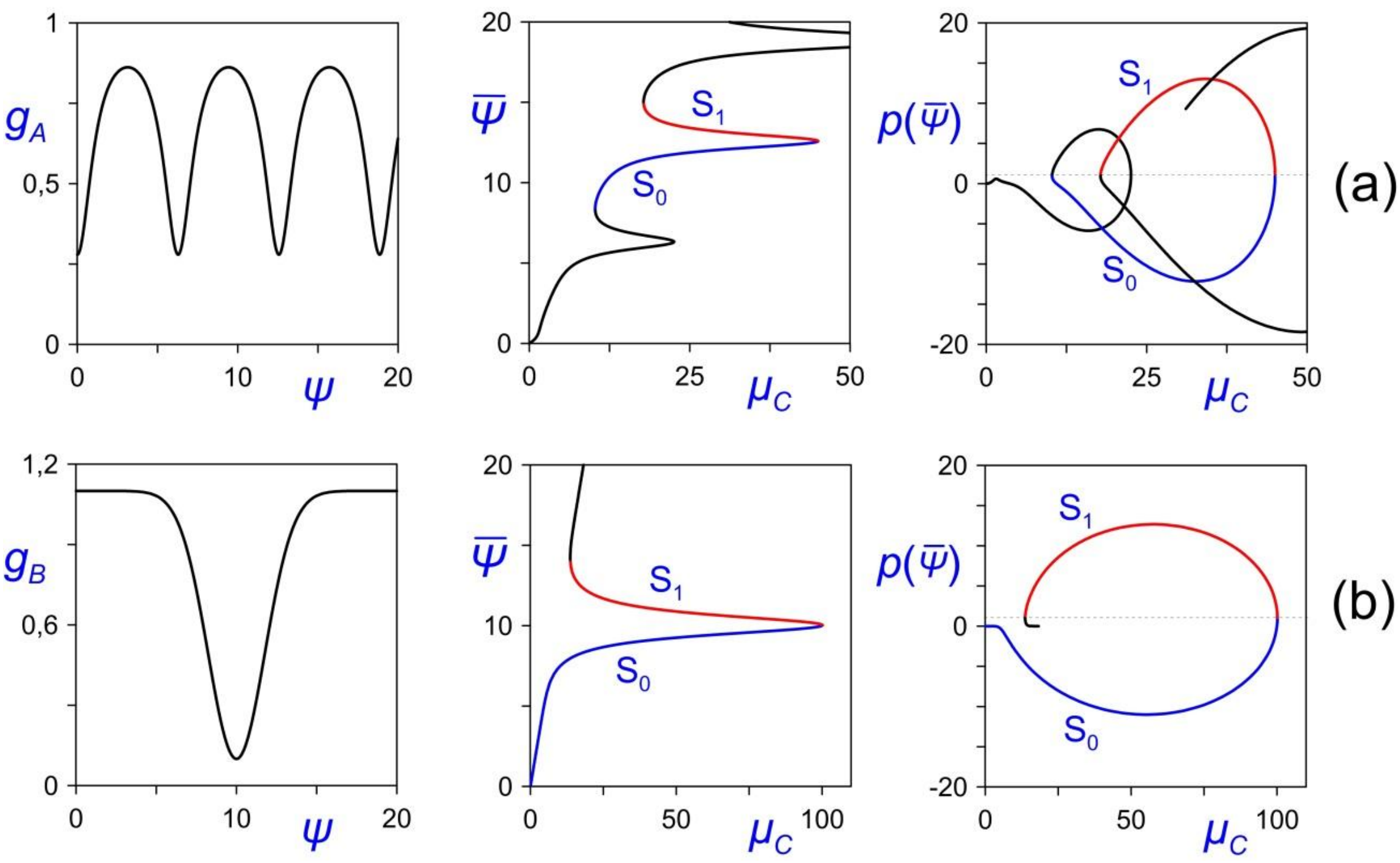

**FIG. A1.** (a) Nonlinear function $g_A(\psi)$, as given by Eq. (A7), corresponding steady-state solution $\overline{\psi}(\mu_C)$ and distribution of $p$ values on such a solution, (b) The same for $g_B(\psi)$, Eq. (A8). $S_1$ and $S_0$ denote the saddle and node fixed points involved in the reported phase-space representations. The horizontal broken line indicates the $p$ =1 value at which the saddle-node bifurcations occur.

Notice also in the set of $(p_j, \omega_j)$ pairs used for the system design that, when ordered according to the frequency, the chosen $p$ values correspond alternatively to the node or the saddle fixed point, i.e., $p < 1$ or $p > 1$. This means that the Hopf bifurcation at $\omega_j$ will occur on the node or on the saddle depending on whether $j$ is odd or even, respectively. On the other hand, the successive $p$ values associated with each fixed point have been chosen such that the successive bifurcations will occur as a function of $\mu_C$ ordered from lower to higher frequency. The ordered alternation between node and saddle [16] is the best choice for the design success since, according to our experience, it assures the stability of the initial fixed point at $\mu_C = 0$ and then the posterior existence of an attractor. Instead the frequency order in the successive bifurcations of a given fixed point is not critical. In general, the properly designed systems have $c_q > 0$, $\forall q$, and $d_q$ of alternatively opposite sign with $d_N = c_N > 0$.

The simple differentiation relation among the variables of system (A2), as given by $z_j = z_N^{(N-j)}$, $j = 1,.., N$, where the superscript denotes the order of time differentiation, implies that the relative presence of the various oscillation modes enhances in proportion to their frequencies when considering variables of successively decreasing subscript $j$ (see Fig. 8 of Ref. [12]). Such a differentiation relation makes also that both the linear part of the vector field of (A2) and the Jacobian matrix are in the companion form. This means that the Jacobian eigenvectors are exclusively determined by the respective eigenvalues as given by $(\lambda^{N-1}, \lambda^{N-2}, .., \lambda, 1)$. In particular, the two-

**TABLE A1.** Coefficients of the systems used to demonstrate the nonlinear oscillatory mixing. The first two columns give the set of chosen ($p_j$, $\omega_j$) values defining the $N$-1 Hopf bifurcations to be imposed in the system design and the third column gives the corresponding time period. The next two columns give the $c_q$ and $d_q$ values of system (A2-A4), which for simplicity have been slightly rounded from the calculated ones, while the next two pairs of columns give the corresponding coefficients of the transformed systems (A11) and (A15), respectively.

| | | | | | **$N = 4$** | | | | |
|---|---|---|---|---|---|---|---|---|---|
| $j$ \ | $p_j$ | $\omega_j$ | $2\pi/\omega_j$ | $c_j$ | $d_j$ | $a_j$ | $b_j$ | $l_j$ | $b_j$ |
| 1 | -4 | 6.283 | 1 | 410 | -82.70 | -4.96 | | 461.75 | |
| 2 | 5 | 41.87 | 0.15 | 22400 | 4280 | 5.23 | -51.75 | -857.19 | -51.75 |
| 3 | -5 | 209.3 | 0.03 | 643000 | -160000 | -4.02 | -37.38 | -127.45 | -37.38 |
| 4 | | | | 312000 | 312000 | 1 | -1.95 | -544.45 | -1.95 |

| | | | | | **$N = 6$** | | | | |
|---|---|---|---|---|---|---|---|---|---|
| $j$ \ | $p_j$ | $\omega_j$ | $2\pi/\omega_j$ | $c_j$ | $d_j$ | $a_j$ | $b_j$ | $l_j$ | $b_j$ |
| 1 | -5 | 0.06283 | 100 | 120 | -18.2 | -6.59 | | 182.09 | |
| 2 | 6 | 0.31416 | 20 | 8335 | 1130 | 7.38 | -62.09 | -309.15 | -62.09 |
| 3 | -6.4 | 2.094 | 3 | 51690 | -8117 | -6.37 | -7.18 | -65.49 | -7.18 |
| 4 | 7 | 20.94 | 0.3 | 32100 | 5659 | 5.67 | -0.70 | -285.17 | -0.70 |
| 5 | -6.6 | 125.664 | 0.05 | 4526 | -869.5 | -5.05 | -0.16 | -90.16 | -0.16 |
| 6 | | | | 39.7 | 39.7 | 1 | -0.044 | -200.29 | -0.044 |

dimensional eigenspace of a Hopf bifurcation with $\lambda_{\pm} = \pm i\omega$ is defined by the vectors

$$\begin{gathered} \left(\ldots,-\omega^{6},0,\omega^{4},0,-\omega^{2},0,1\right), \\ \left(\ldots,\omega^{5},0,-\omega^{3},0,\omega,0\right), \end{gathered} \tag{A9}$$

so that the various limit cycles of frequencies $\omega_j$ will appear accordingly oriented, independently from which fixed point they arise. Of course these peculiarities would become deeply hidden in a system transformation generically producing new variables as arbitrary combinations of those of the standard system (A2).

When trying to identify what features of the system of equations are responsible for the good working of the oscillatory scenario, one should look for the three kinds of ingredients that seem needed for it: feedback, nonlinearity, and competition on the ensemble of interrelations among the variables and their time rates of change. Feedback is clearly manifested in the circuits of influences, always connected in a closed structure due to the coupling of the equations. Nevertheless, its analysis loses sense by realizing that arbitrary coordinate transformations can yield new sets of variables sustaining deeply different circuits of interrelations while the behaviors remain qualitatively equivalent. Nonlinearity is clearly identified in the functional dependences of the interrelations and in the case of system (A2-A4) it is exclusively contained in $g(\psi)$, being its primary role to allow for the existence of the saddle-node pair of fixed points [17]. More difficult is to appreciate where the occurrence of successive Hopf instabilities on the fixed points resides. The oscillatory behavior should be based on competing effects in the relational circuits of influences and large numbers of coexisting oscillation modes require a well-organized structure of competition. In the case of a properly designed system, it seems to be associated with the alternatively opposite signs of the $d_q$ coefficients, which, according to Eq. (A4), imply that the variations of two successive variables affect $\psi$ in opposite sense while, at the same time, one of the variables determines the time rate of change of the other, and this for each pair of successive variables. At this respect, there is the curious fact that the coefficients of the designed systems satisfy an approximate relation with the $p$ values at which the fixed points should bifurcate. Having chosen $c_N = d_N$, such a relation is

$$p_j \approx \frac{c_{N-j}}{d_{N-j}},\ j = 0,1,..,N-1, \tag{A10}$$

where $p_0 = 1$ corresponds to the saddle-node bifurcation and the rest of $p_j$ to the Hopf bifurcations at $\omega_j$, with $j = 1, 2, .., N\text{-}1$ and the frequencies always ordered from lower to higher. The more different the $\omega_j$ the more approximated relation (A10) becomes and it is worth noting that the relation applies also for nonproperly designed systems.

The various variables and coefficients of Eqs. (A2-A4) have different time dimension and, to facilitate the association with the various timescales, it is useful to transform to time dimensionless coordinates like $y_j = d_j z_j$, provided the $d_j$ are different from 0. The system then becomes

$$\begin{aligned} \frac{dy_1}{dt} &= d_1\left[-\sum_{q=1}^{N} a_q y_q + \mu_C\, g\left(\psi,\mu\right)\right], \\ \frac{dy_j}{dt} &= b_j y_{j-1},\ j = 2,..,N, \end{aligned} \tag{A11}$$

with

$$\psi = \sum_{q=1}^{N} y_q, \tag{A12}$$

$$a_q = \frac{c_q}{d_q},\ q = 1,2,..,N, \tag{A13}$$

$$b_q = \frac{d_q}{d_{q-1}},\ q = 2,3,..,N, \tag{A14}$$

and where only $d_1$ and $b_q$, $q = 2, 3, .., N,$ account for time dimensionality. In properly designed systems, $d_1$ is either negative or positive depending on whether $N$ is even or odd, respectively. The $d_q$ sign alternation is now transferred to the $a_q$ coefficients that, according to relation (A10), are approximately given by the $p$ values of the Hopf and saddle-node bifurcations as $a_q \approx p_{N-q}$. Notice in Eq. (A11) the ordered correlation in the terms $a_q y_q \approx p_{N-q} y_q$ between the frequency sensitivity of $y_q$ and the Hopf frequency associated with $p_{N-q}$, for $q$ = 1, 2, .., N-1, i.e., $y_1$ is the fastest variable while $p_{N-1}$ corresponds to the fastest Hopf frequency and so on. The $d_q$ sign alternation implies also that $b_q < 0, \forall q$, and these negative values reflect the sequence of dynamical competition between successive pairs of variables, i.e., a variation of $y_j$ negatively influences the time rate of change of $y_{j+1}$ and this successively from $j$ = 1 to $j$ = $N$-1. At the same time, by determining the magnitudes of such a sequence of influences, the $b_q$ values characterize in some way the multiplicity of timescales. Nevertheless, the actual oscillation frequencies depend on the rest of parameters also and it is difficult to find a definite, although rough, relation among the $b_q$ and the $\omega_j$.

It remains the fact that, according to Eq. (A12), all the variables participate in the argument of the nonlinear function $g(\psi)$ and it is unclear to what extent this is compulsory for the oscillatory scenario achievement. Such a condition may be relaxed by introducing $\psi$ as one of the variables instead of, for instance, $y_1$, so that the system becomes

$$\frac{d\xi_1}{dt} = -\sum_{q=1}^{N} h_q \xi_q + d_1 \mu_C \, g\left(\xi_1, \mu\right),$$

$$\frac{d\xi_2}{dt} = b_2 \left( \xi_1 - \sum_{q=2}^{N} \xi_q \right), \tag{A15}$$

$$\frac{d\xi_j}{dt} = b_j \xi_{j-1} \; , \; j = 3, 4, .., N \; ,$$

with $\xi_1 = \psi$ and $\xi_{j \neq 1} = y_j$, and where

$$h_1 = d_1 a_1 - b_2,$$

$$h_q = d_1 a_q - b_{q+1} - h_1 \; , \; q = 2, 3, .., N-1, \tag{A16}$$

$$h_N = d_1 a_N - h_1.$$

and the $b_q$ as defined in Eq. (A14). Of course, one can imagine further coordinate changes that, by maintaining $\psi$ as one of the variables, transform the system to the generic form of Eq. (A1) with, of course $m$ =1, and the matrix $A$ and vector $b_1$ both full of non-vanishing coefficients, but with the nonlinear function remaining a function of one of the variables only, as it should be even given by Eq. (A3). Thus, the participation of the various variables into the argument of the nonlinear function is not a necessary condition for the oscillatory development while the required interrelations among variables can work through the linear part of the vector field and the appropriate presence of the nonlinear function into the several equations.

A final query concerns the analysis of to what extent the considered system admits to be transformed so that it could be seen like a set of coupled subsystems, with each one of them sustaining some of the $N$-1 oscillation modes when decoupled from the others. Or, more in general, to what extent the generalized Landau scenario is achievable by combining a number of single-frequency oscillators through appropriate coupling. The difficulty for this purpose arises from the required participation of two fixed points, the saddle-node pair, and the occurrence of Hopf bifurcations on both, while the coupling of $n$ oscillators usually produces a $2n$-dimensional system sustaining $n$ oscillation modes around a single fixed point and with their combination related to the linear superposition of oscillations occurring for very weak coupling.